\documentclass[11pt]{article}
\usepackage{epsf}
\begin{document}
\def\dx{\partial_x}
\def\deltamn{\delta_{m+n,0}}
\def\deltaxy{\delta(x-y)}
\def\levi{\epsilon_{ijk}}
\def\crossa{/ \hspace{-.4cm} \nwarrow}
\def\crossb{\setminus \hspace{-.4cm} \swarrow}
\def\crossc{=}
\def\crossd{|\;|}
\def\rlx{\relax\leavevmode}
\def\inbar{\vrule height1.5ex width.4pt depth0pt}
\def\IZ{\rlx\hbox{\small \sf Z\kern-.4em Z}}
\def\IR{\rlx\hbox{\rm I\kern-.18em R}}
\def\ID{\rlx\hbox{\rm I\kern-.18em D}}
\def\IC{\rlx\hbox{\,$\inbar\kern-.3em{\rm C}$}}
\def\IN{\rlx\hbox{\rm I\kern-.18em N}}
\def\one{\hbox{{1}\kern-.25em\hbox{l}}}
\def\smallfrac#1#2{\mbox{\small $\frac{#1}{#2}$}}

\begin{titlepage}

December, 1995 \hfill{UTAS-PHYS-96-03}\\
\mbox{}\hfill{hep-th/yymmdd}
\vskip 1.6in
\begin{center}
{\Large {\bf  Chord diagrams and BPHZ subtractions}}
\\[5pt]
\end{center}

\normalsize
\vskip .4in

\begin{center}
Ioannis Tsohantjis
\hspace{3pt} Alex C Kalloniatis\footnote{
Institute for Theoretical Physics III,
University of Erlangen-Nuremberg, D-91058 Erlangen, 
Germany}
 \hspace{3pt} and Peter D Jarvis\footnote{
Alexander von Humboldt Fellow ~~
email: \emph{Peter.Jarvis@phys.utas.edu.au}}

\par \vskip .1in \noindent
{\it Department of Physics, University of Tasmania}\\
{\it GPO Box 252C Hobart, Australia 7001}
\end{center}
\par \vskip .3in

\begin{center}
{\Large {\bf Abstract}}\\
\end{center}

\vspace{1cm}
The combinatorics of the BPHZ subtraction scheme for a
class of ladder graphs for the three point vertex in $\phi^3$
theory is transcribed into certain connectivity 
relations for marked chord diagrams (knots with transversal intersections).
The resolution of the singular crossings using the equivalence relations
in these examples provides confirmation of a proposed fundamental 
relationship between knot theory and renormalization 
in perturbative quantum field theory.
\end{titlepage}

\section{Introduction and Discussion}
Despite the predictive successes of quantum field theory established 
over the last half century since the formulation of the renormalization
prescription, the mathematical well-foundedness of the
procedure has been questioned since its very inception. As is well known, 
alternative approaches such as axiomatic, $S$-matrix and ultimately 
string theories have been spawned as a result. On the other hand, 
recent study of low-dimensional quantum field theory ($D=2$ or 3) 
has established deep connections with mathematics and geometry. 
In this note we consider whether such developments can be related to the 
traditional renormalization programme of perturbative quantum field theory 
($p$QFT) in arbitrary dimensions. Recently 
Kreimer\cite{Kreimer1,Kreimer2,Kreimer3} and
Kreimer and Broadhurst\cite{KreimerB1} have proposed a 
relationship between Feynman graphs in $p$QFT and knot theory,
developed in terms of skein relations on link diagrams
associated with the Feynman graphs. Here we 
attempt to provide an alternative concrete formulation,
as well as formal confirmation, of these ideas.
Specifically, for a  class of ladder-type vertex corrections in $\phi^3$ 
theory, we show how the graphs' contributions to 
the renormalization constant, including the correct counterterm 
subtractions, can be obtained by using equivalence relations on 
associated chord and singular knot diagrams.

The class of Feynman graphs considered here is that of  
vertex corrections in the ladder approximation,
in which the rungs are simple or crossed propagators. 
Below, we firstly note (\S 2) that such graphs have a direct interpretation in 
terms of marked chord diagrams\cite{Alvarez,Bar-Natan,Birman}. Each such 
diagram can be mapped to a knotted loop, including
transversal intersection points\cite{Bar-Natan, Kauffman}.
Then, motivated by the combinatorics of the 
BPHZ\cite{BPHZ} subtraction scheme, and features of the types of divergent
integrals which arise for the graphs\cite{Kreimer1,Kreimer2,Kreimer3},
we formulate equivalence relations between such diagrams.
The result of the iterative application of the rules is an expression for 
the value of the divergent part of each graph, plus its counterterm 
subtractions, which gives the graph's contribution to the renormalisation 
constant at the appropriate order. The procedure is illustrated
with several examples (\S 3). These clarify the
suggestion\cite{Kreimer1,Kreimer2,Kreimer3,KreimerB1}
that certain types of Feynman graph `topology', contributing to
the renormalization constant with transcendental number coefficients,
should be associated with knots. Finally, a more abstract, algebraic
setting for the transcription of Feynman graphs into the language of 
knots is given (\S 4). In our concluding remarks we suggest how the
proposed formalism can be extended to arbitrary Feynman graphs and divergences,
and speculate on the implications of the work for the significance of the 
renormalization procedure in $p$QFT. 
   
\section{Marked chord diagrams and counterterm subtractions}
In the BPHZ approach\cite{BPHZ} to renormalization, the regularization
of a given Feynman graph $\gamma$ is specified by 
of a sum of subtracted terms $\gamma \! \setminus \! \Gamma \cdot 
\prod_{\alpha \in \Gamma} {\cal T}(\alpha)$ 
generated by disjoint families 
$\Gamma = \{\alpha_1, \alpha_2, \ldots \}$
of divergent subgraphs. The explicit construction
involves specific operations on the respective integrands
(Taylor expansions with respect to external momenta); nevertheless the 
combinatorics should be seen in terms of operations on the graph $\gamma$ 
as a whole (indeed, the additional terms correspond to the counterterm 
subtractions in the usual formulation involving renormalization constants).
It is at this level that the transcription into the language of manipulating
singular knot diagrams should be seen.

We deal with $\phi^3$ theory, and for the present consider only
corrections to the basic one loop vertex graph in the ladder 
approximation, where only the internal rungs
(propagators) may be crossed, and no self-energy graphs,
or dressings of internal propagators or vertices, are allowed.
The formalism to be developed requires for its starting point
that the Feynman graph be representable as a chord 
diagram\cite{Alvarez,Bar-Natan,Birman}. For this
class of ladder corrections the transcription is immediate:
the original loop serves as the circle on which all vertices are located, 
and the internal propagators are viewed as chords (which may cross).
In order to retain knowledge of the external vertices
these are marked with ticks on the appropriate external
arcs. Examples of Feynman graphs and their chord transcriptions, together
with some graphs which are not admitted for present purposes, are given in 
Figure \ref{chordexx}.
Obviously, by restricting ourselves to this class of diagrams,
we avoid issues of two-point and overlapping divergences. It should be 
emphasized that the discussion is intended simply to illustrate the 
working assumption that it is useful, at least for renormalization 
purposes, to associate Feynman graphs to knots -- we do \textit{not} attempt a 
derivation of this association. However, for a more abstract setting for 
the formalism, and consideration of possible
extensions to arbitrary Feynman graphs, see the concluding remarks below.

Having arrived at a chord characterisation $D_\gamma$ of the Feynman graph
$\gamma$, we associate it\cite{Bar-Natan} with an oriented singular (framed) knot 
$K_\gamma$, where the singularities correspond to transverse
self-intersection points, represented as filled nodes. This is done by 
traversing the circle in a given sense (say, anti-clockwise), and 
associating with each chord a crossing point, so that the two ends of 
each chord are identified. Since these are the only allowed self-intersections, 
the necessity for these to be consistent will generally 
force a choice at several other places as to which arcs over or undercross,
so that there is in principle a family of possibilities
(marks are transferred to the appropriate arcs).
Some examples are given in Figure \ref{knotexx}, together with an
extended notation for the nonstandard examples wherein marked crossings
are introduced and represented by unfilled nodes. 

\begin{figure}[tbp]
\epsfxsize=4in
\centerline{\epsfbox[0 0 771 254]{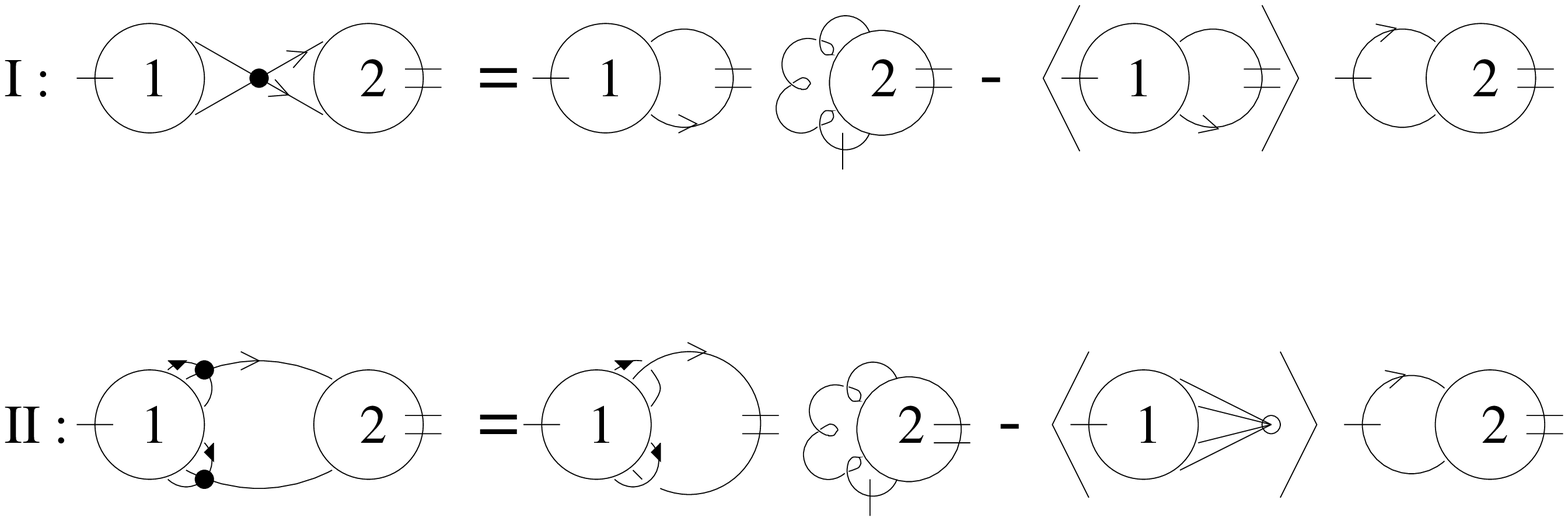}}
     \vspace{8cm}
	\caption{Rules $I$, $I\!I$ for manipulating singular knot diagrams derived 
	from chords. The increase of framing number on the first `2' term on the 
	right hand side 
   is the number of nodes of the `1' part (including the explicitly 
   drawn ones on the left hand side).  For 
	discussion of the unfilled nodes appearing in $I\!I$, see the text and the 
	examples given in Figure \ref{equivexx}.}
	\protect\label{equivreln}
\end{figure}

The basic rules by which we propose the knot diagrams are to be manipulated 
are the 
connectivity relations $I,I\!I$, described in 
Figure \ref{equivreln}. Here `1', `2' are understood to be parts 
which remain unchanged wherever they appear, 
whose connections with the remainder 
occur smoothly at the boundaries. The connectivity properties allow the 
singularities to be resolved in terms of conventional over- and under- 
crossings. The standard discussion\cite{Bar-Natan} 
\emph{disallows} isolated chords (so that the 
left-hand side of $I$ would vanish). Here however it is precisely the 
more involved relations $I$, $I\!I$ which generate the counterterm 
subtractions. Both ensure that two-line reducible knots can be simplified 
in terms of indecomposable parts (reflecting the subdivergence structure 
of the original graph). An additional essential notion is that of 
`bracketing' $< \cdot >$, which is allowed to any (finite) recursive depth
(see below). Also, each of the disconnected parts `1', `2' inherits the
other's additional markings, in order that it be 
regarded as a legitimate three-point contribution (the distinction
between `1' and `2' is determined by which side carries the double marking
(this may be distributed between different arcs or propagators as in the 
last examples of Figures \ref{chordexx}, \ref{knotexx})).

Each of $I$, $I\!I$ involves an \emph{unbracketed} term on the right hand side,
in which the transversal crossings are either \emph{dis}solved by
smoothing (as in $I$) or \emph{re}solved by specifying over- and 
under-crossings ($I\!I$). On the `2' term there appear additional twists 
on the closing loop, corresponding to an increase of framing number (in the 
blackboard framing) equal to the grading of `1' (the total 
number of nodes on the `1' side, including the explicitly drawn ones).
Further, in the bracketed term of $I\!I$, `1' is closed not by an 
additional loop, but by a marked vertex (represented by an unfilled node)
formed by fusing the original 
transversal crossings, with the elimination of two of the original six 
connecting lines to the remainder (in the same way, $I$ can be thought of as 
a special case of $I\!I$ in which the two crossings are fused, with the small
connecting arcs cancelling each other out). Which lines survive appears to depend
on the details of these connections, but can be 
established for particular examples (see \S 3 below).
Finally, it will be apparent from these examples that
the rules must be applied in a particular order to the diagrams'
two-line reducible locations (working left to right as drawn in the figures),
and further that the bracket on the right-hand side embraces
not only the `1' part but also any additional disconnected terms
which multiply the expression on the left-hand side. 
The necessity to refer back to the original BPHZ combinatorics for these 
points can be expected to be overcome in a more complete treatment
(see concluding remarks below).

The result of the application of the rules to a given knot
$K_\gamma$ is an expression involving products of bracketed and 
unbracketed framed knots. Following 
Kreimer\cite{Kreimer1,Kreimer2,Kreimer3,KreimerB1},
the central claim is that the divergent contribution of the original Feynman 
diagram $\gamma$, including counterterm subtractions if any, has precisely this 
structure, where there is a systematic association between the component 
parts and certain basic divergent integrals. The (iterated) brackets signal
taking divergent parts (see (\ref{bracketeqn}) below) with respect to the 
regularization parameter, and the projection on to the divergent piece of the 
overall result is the graph's contribution to the renormalization 
constant. (Strictly speaking, the renormalization
point, and the precise dependence on momentum scale, should be spelled out. For
the present sketch of the combinatorics we 
work with a `minimal subtraction' type scheme\cite{Kreimer1},
but the details are not required).

\section{Examples}  

\begin{figure}[tbp]
\epsfxsize=4in
\centerline{\epsfbox[0 0 684 363]{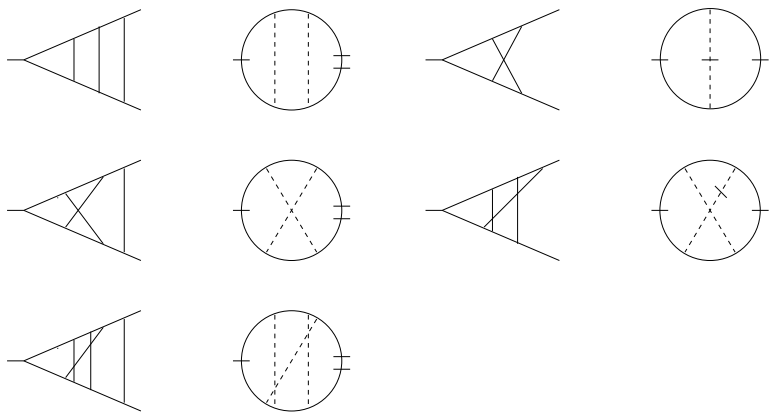}}
     \vspace{8cm}
 	\caption{Examples of diagrams and their chord transcriptions. The last 
 	two are given by an extended notation in which a single \emph{internal} 
 	propagator is marked.}
 	\protect\label{chordexx}
 \end{figure}

\begin{figure}[tbp]
\epsfxsize=4in
\centerline{\epsfbox[0 0 680 420]{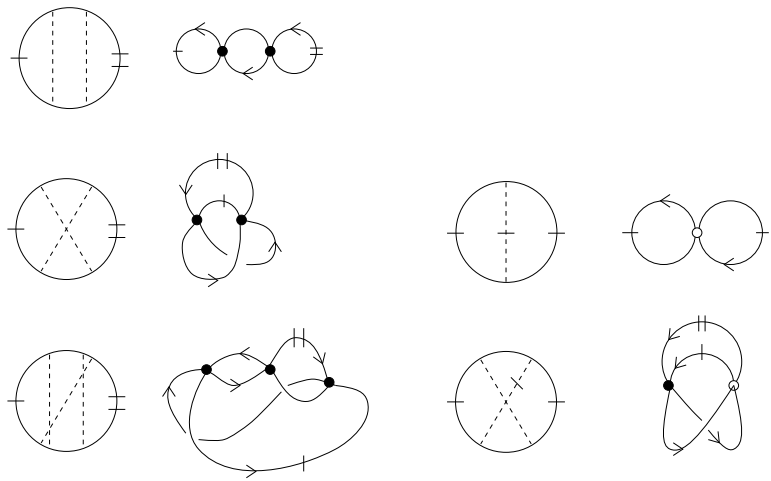}}
    \vspace{8cm}
	\caption{Examples of knots corresponding to the chords of Figure 
	\ref{chordexx}. The last two are 
	handled by introducing unfilled nodes for the relevant knot crossings.}  
	\protect\label{knotexx}
\end{figure}

\begin{figure}[tbp]
\epsfxsize=5in
\centerline{\epsfbox[0 0 731 520]{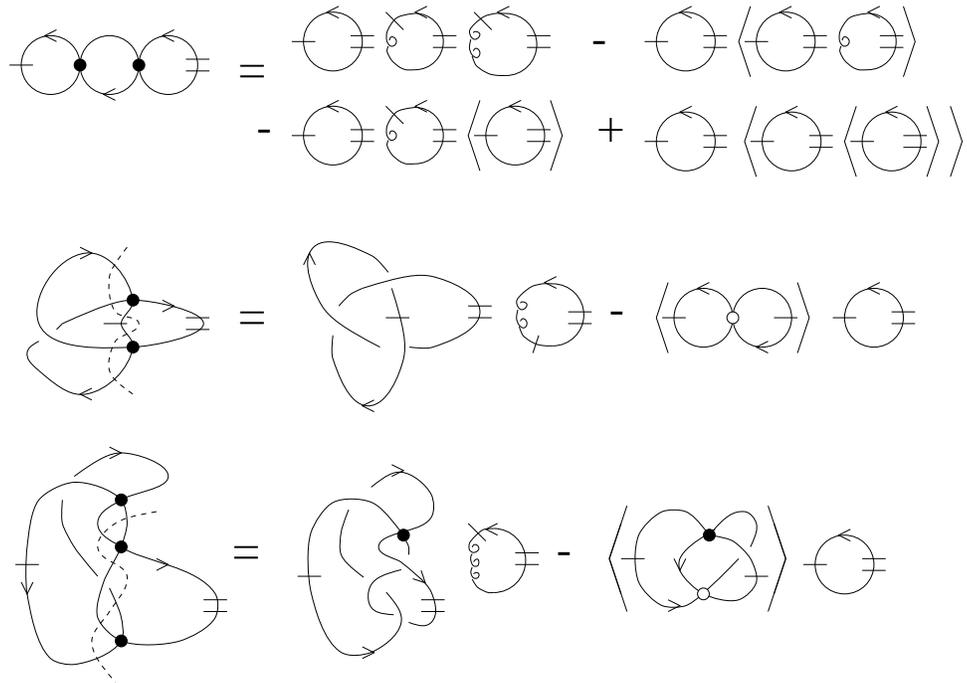}}
     \vspace{10cm}
	\caption{Examples of rules for manipulating knot diagrams derived from 
	chords. In these cases the connections of the unfilled nodes are given 
	explicitly (see Figure \ref{equivreln})}
	\protect\label{equivexx}
\end{figure}

We illustrate the technique with some examples (see Figure \ref{equivexx}) before 
giving a more abstract discussion. The first case in the figure shows the 
result for the case of simple ladder corrections to the vertex 
(corresponding to the first entry of Figure \ref{chordexx} and \ref{knotexx})
at three loops; higher loops follow in the same way. Obviously only relation $I$
need be applied, and the result is a sum of terms involving bracketed and 
unbracketed products of simple unknotted loops with framing number increasing 
up to two. This accords \emph{precisely} with the algebraic 
contribution of this graph to the renormalization constant when the 
subtractions are taken into account. The basic three loop graph is accompanied
in the sum over disjoint families of divergent subgraphs
($\gamma \! \setminus \! \Gamma$) by the one and two 
loop counterterms ($\prod_{\alpha \in \Gamma} {\cal T}(\alpha)$) corresponding
to the bracketed pieces in the figure. Only massless propagators
contribute (infrared divergences being avoided in the presence of the
subtractions), and as pointed out in \cite{Kreimer1}, at least in dimensional 
regularization, the multiloop divergent integrals possess an iterative kernel structure
whereby the key ingredient is
\begin{equation}
	\mbox{}_k\Delta =
	(q^2)^{(k+1)\varepsilon} \int d^d p \frac{(p^2)^{- k\varepsilon}}{p^4(p+q)^2}
	\label{Deltadefn}
\end{equation}
which by Lorentz invariance is independent of the direction of $q$
(note that we are working in $\phi^3_{D=6}$). The result is a product of 
such terms with rising values of $k$, represented graphically by the 
increase of framing (or writhe number in the blackboard framing) in the figure.     
Finally, the contribution of the graph to the renormalization constant,
in direct correspondence with the diagrammatical terms, is
\begin{equation}
	<Z> = \left< \mbox{}_0 \Delta \;\mbox{}_1 \Delta \;\mbox{}_2 \Delta
	\, - \, \mbox{}_0 \Delta <\mbox{}_0 \Delta\; \mbox{}_1 \Delta> \, - \,
	\mbox{}_0 \Delta\; \mbox{}_1 \Delta <\mbox{}_0 \Delta> \, + \,
	\mbox{}_0 \Delta <\mbox{}_0 \Delta <\mbox{}_0 \Delta>> \right>,
	\label{3loopladderZ}
\end{equation}
where now $< \cdot >$ signifies taking the divergent part 
(in minimal subtraction).

The second and third examples in Figure \ref{equivexx} correspond to the 
second and third entries in 
Figures \ref{chordexx} and \ref{knotexx}. In each case there is only a single
divergent subgraph, so only two terms on the right hand sides appear. 
Now, however, the connectivity relation $I\!I$ must be applied
(as indicated by the dotted shadings on the left hand sides).
The bracketed terms specify the correct counterterms, namely the 
divergent pieces corresponding to the primitive two loop Feynman graphs
with two and three crossed propagators (compare the fourth 
and fifth entries in Figures \ref{chordexx} 
and \ref{knotexx}). Concerning the point mentioned
above about the connections of the marked (unfilled) node to the rest of the 
diagram in the bracketed `1' term of the equivalence relation $I\!I$,
it can be seen here how the rule is to be interpreted. Consider the path 
starting with the upper node, continuing to the lower node via what would have 
been the route through `2' (but now smoothed), and returning to 
to the upper node within `1'. From these cases it appears that this entire
line segment is eliminated, with the connections to the rest of the diagram
being taken up by the replacement marked (unfilled) node.
The markings are here given so as to agree with
the needed counterterms, but with the elimination of the said line segment,
it is not apparent what the general rule for reassigning the second mark might 
be. As emphasized at the outset, the answers arrived at here arise by explicit 
reference back to the BPHZ algorithm for these cases:
in the absence of a transcription of \emph{arbitrary}
Feynman graphs into chord diagrams, the rule cannot be given
in complete generality (see concluding remarks below about extensions of 
the present approach). 
 
Let us examine the unbracketed terms in these examples.
In the first case, there are only two singular crossings to be resolved, 
and there emerges for the `1' term a standard knot, a marked version of the 
trefoil (multiplied by the simple unknotted loop with the correct framing number).
In the second case, one singular crossing remains in the `1' term.
Clearly the additional entanglements 
engendered by $I\!I$ prevent the further application
of $I$ to this remaining node. In general, the only possibilities for
further simplification using $I$ or $I\!I$ should be where there is 
genuine two-line reducibility arising from other parts of the original 
Feynman graph.

In the case of the simple ladder graphs, the diagrammatic expression
after applying $I$, $I\!I$ was noted above to be in direct 
correspondence with the algebraic structure of the contributions to the 
renormalization constant, and a mapping could be established
between framed unknots and divergent one loop integrals,
in agreement with \cite{Kreimer1,Kreimer2,Kreimer3,KreimerB1}. 
For more complicated `topologies', such as the second and third examples 
in Figure \ref{equivexx} (via Figures \ref{chordexx}, \ref{knotexx}) 
which we are considering, divergent integrals of 
increasing complexity are obviously involved. The central claim of 
\cite{Kreimer1,Kreimer2,Kreimer3,KreimerB1} is that there 
is in all cases a systematic association between the algebraic structure 
and the results of diagrammatic manipulations 
(developed there in terms of skein relations on link diagrams
associated with the Feynman graphs).
The ubiquitous appearance of for example 
$\zeta(3)$ in certain three loop integrals is thus to be interpreted as 
a signal of the repeated occurrence of a particular diagrammatic term,
claimed in fact to be the trefoil, with higher transcendentals coming with 
other knots (the simple ladder graphs, associated with unknots,
being free of transcendentals of this 
type\cite{Kreimer1,Kreimer2,Kreimer3,KreimerB1}).

\begin{figure}[tbp]
\epsfxsize=5in
\centerline{\epsfbox[0 0 547 74]{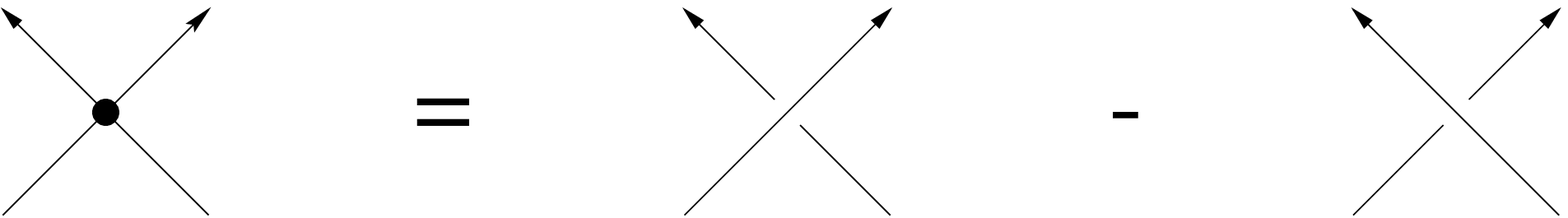}}
\vspace{6cm}
	\caption{The Vassiliev property $V$ expressing a singular crossing as a 
	difference between conventional over- and under- crossings. The complete
   set of equivalence relations would involve $I$, $I\!I$, together with new 
   relations $I\!I\!I$, $I\!V$ covering two point divergences, as well as $V$.
   $V$ is used in the text in the interpretation of Figure \ref{equivexx}.} 
	\protect\label{Vassequiv}
\end{figure}

The second and third examples 
in Figure \ref{equivexx} which we are considering provide 
confirmation of these claims, from the viewpoint of the present approach.
In fact, as noted already, the second example yields directly the trefoil, 
in accord with the occurrence of $\zeta(3)$ in the leading divergence.
In the third example, however, not all the singular knot crossings are 
resolved, and this will certainly be the case in general.
At this stage it should be pointed out that we have until now 
made no use of an additional property which is
usually applied\cite{Bar-Natan}, namely the Vassiliev type 
equivalence relation, whereby singularities 
are resolved in terms of a difference between standard over- and under- crossings
(given as $V$ in Figure \ref{Vassequiv}). In fact this relation is 
already implicit in the translation between chord and knot diagrams:
provided that one is evaluating a singular knot with a Vassiliev invariant of 
type $n$, it is immaterial whether under- \emph{or} over-crossings are used in 
the specification of the avoided crossings (because the difference is a 
knot with $n\!+\!1$ singularities, on which the Vassiliev invariant vanishes).
From this, we would expect $V$ to be involved in the detailed 
interpretation of our formalism; again, it should be borne in mind that 
we are building only on the limited examples at hand. For the moment, we 
can make use of this idea in interpreting the third example, and note that
one part of the unbracketed knot under $V$ is certainly a version of 
the trefoil (possibly in a different projection, and with different marking 
on the arcs from the previous case). Similarly, the bracketed part 
(corresponding to the primitive divergence, the fifth entry in Figures 
\ref{chordexx} and \ref{knotexx}) also has the `topology' of the trefoil (in fact 
the corresponding $p$ loop Feynman graph with one propagator crossing
successive rungs has been associated\cite{Kreimer1,Kreimer2,Kreimer3,KreimerB1}
with the $(2,2p\!-\!3)$ torus knot, with transcendental contributions 
involving $\zeta(2p\!-\!3)$).

\section{Abstract setting} 
 
Up to now, we have presented a diagrammatic recipe in order to reproduce the 
combinatorics of the BPHZ subtraction scheme for a particular class of 
Feynman graphs. However, an algebraic setting ({\it c.f.}
\cite{Bar-Natan,Turaev,Turaevtop}) can be given which stands alone, 
and hopefully is general enough to encompass all graphs of the theory
(see concluding remarks below). Firstly introduce the free module $\cal D$ of
marked chord diagrams, given the structure of an abelian algebra by 
juxtaposition. The `scalars' of this module comprise the ring $\Lambda$
of (one-sided) Laurent series in the regularization parameter
(the cutoff momentum scale, or the $\varepsilon$ parameter in dimensional 
regularization with $d=D-2\varepsilon$). Inherent in this ring is the 
operation of the projection onto the divergent part: if $\lambda \in 
\Lambda$ we have
\begin{eqnarray}
	\lambda &=& \frac{\lambda_{-N}}{\varepsilon^N} + 
	\frac{\lambda_{-N+1}}{\varepsilon^{N-1}}
	+  \cdots + \lambda_0 + \lambda_1 \varepsilon^1 + \cdots, \nonumber \\
	<\lambda>&=& \frac{\lambda_{-N}}{\varepsilon^N} +
	\frac{\lambda_{-N+1}}{\varepsilon^{N-1}}
	+  \cdots + \lambda_{-1}.  
	\label{bracketeqn}
\end{eqnarray}
Correspondingly, the notion of \emph{bracketing} is introduced as an abstract 
operation in $\cal D$ and in the associated algebra $\cal A$ of 
singular knots. The  
connectivity relations $I, I\!I$ generate an ideal $\cal I$ in $\cal A$. 
The implementation of the equivalence relations amounts to working in the factor 
algebra ${\cal P} = {\cal A}/{\cal I}$. ${\cal P}$ is generated as a 
(polynomial) algebra in the indecomposable parts and their bracketings, 
which consist of marked, framed knots $K_i$ (possibly including unresolved 
nodes; see examples and concluding remarks below). Thus, for the knot 
$K_\gamma$ associated with the chord diagram $D_\gamma$ for the Feynman 
graph $\gamma$, we have
\begin{equation}
	K_\gamma = \sum_{(i)(j)(k)\cdots}c_{(i)(j)(k)\cdots}K_{i_1}K_{i_2} \cdots
	<K_{j_1} K_{j_2} \cdots<K_{k_1}\cdots<\cdots >> \cdots>.
	\label{knotequivclasses}
\end{equation}
Finally, the renormalization constant contribution should be seen as the 
divergent piece of a homomorphism $Z$ from ${\cal P}$ to $\Lambda$
for which $Z(K<K'>) = Z(K)<Z(K')>$:
\begin{equation}
	Z_\gamma = <Z(K_\gamma)> = \left<\; \sum_{(i)(j)(k)}c_{(i)(j)(k)}
   \Delta_{i_1}\Delta_{i_2} \cdots
	<\Delta_{j_1} \Delta_{j_2} \cdots<\Delta_{k_1}\cdots<\cdots >> \cdots>
    \; \right>
	\label{finalZ}
\end{equation}
where $Z(K_i) = \Delta_i$ are basic scalars (Laurent series) 
fixed once and for all in the theory (see the foregoing discussion in
connection with the examples given above).

\section{Conclusions}

In this note we have exploited the possibility of interpreting a class 
of ladder diagrams in $\phi^3$ theory directly as chord diagrams, in order 
to present the BPHZ subtraction scheme in terms of equivalence relations on 
associated singular knot diagrams. The resulting structure is found to be present 
in the evaluation of the renormalization constant, if the various 
diagrammatic terms are taken to stand for algebraic contributions from
basic divergent integrals. Moreover, the known association between 
certain diagram `topologies' and transcendental number contributions
to the renormalization constant (for example the occurrence of $\zeta(3)$ at 
three loops) is made more definite, in that the 
contributions from `topologies' corresponding to
indecomposable parts in the knot language, and as confirmed
in the examples discussed, can in fact be associated 
with specific knots ($\zeta(3)$ with the trefoil, or $\zeta(2p\!-\!3)$ with the 
$(2,2p\!-\!3)$ torus knot\cite{Kreimer1,Kreimer2,Kreimer3,KreimerB1}).

Before speculating on the significance of the view
of renormalization suggested in \cite{Kreimer1,Kreimer2,Kreimer3, 
KreimerB1} and, supported and advocated more concretely by the present work,
let us emphasize two major limitations of our
approach. Firstly we have \emph{only} been 
concerned with a simple class of ladder corrections to the three-point 
vertex, for which the chord transcription is immediate. In Figure \ref{knotexx}
are given examples which do not comply -- it is clear that in general, chords
incorporating trivalent vertices are needed. All our conclusions are
thus predicated on the assumption that moves can be 
introduced\cite{Bar-Natan} whereby these more general cases can be reduced to
the present framework. Secondly, we have of course ignored all 
considerations of two-point functions, and associated problems of 
overlapping divergences. In a sense, this second type of extension of the formalism
poses less of a problem than the first, in that it can be envisaged how 
appropriate new equivalence relations $I\!I\!I$, $I\!V$ might be introduced to 
cover connectivity between two parts, each of which carries only \emph{one} 
external marked line. Further issues, such as \emph{form factors} 
(for example the distinction between mass and wave-function renormalizations
in the two-point function), can be handled by a matrix 
formalism\cite{Kreimer1}. Work to handle these two major limitations,
in order to give a complete picture of $\phi^3$ theory, is
currently in progress.

Subject to these assumptions about the general validity of
our work, the picture of $p$QFT afforded by our current approach is 
the following. Associated with each renormalizable theory are certain 
fundamental invariants $Z$ which, technically, belong to the algebraic dual
of the algebra $\cal P$ of indecomposables introduced above
(in the case of form factors they may be matrix valued over $\Lambda$).
The contribution of each Feynman graph to the appropriate
renormalization constant is then the infinite part of the evaluation of
the corresponding $Z$ on the associated $K_\gamma$ as in (\ref{finalZ})
above, $Z_\gamma =\;<\!Z(K_\gamma)\!>$ (or a component thereof in the case of 
form factors). Thus in order to evaluate $Z_\gamma$ it is in principle 
only a matter of expressing $K_\gamma$ in terms of equivalence classes in 
$\cal P$, and applying the homomorphism $Z$ assuming that its action is known
on an algebraic basis for $\cal P$.
Moreover, one may suggest further that the  
renormalization constant as a whole to all orders
in $\hbar$, involving a summation over \emph{all} Feynman graphs, 
has the significance of a true topological (manifold) invariant 
(although we do not speculate at this stage on the precise 
nature of such an invariant). This status would then accrue 
also to many of the standard functions of $p$QFT which depend on the 
renormalization process for their definition, such as the $\beta$ 
functions\cite{BDK}. 
Of course, the characterization of the algebra $\cal P$ and its dual,
and the construction of topological invariants therefrom
may be a difficult mathematical problem (compare discussions of Conway 
algebras and skein quantization for arbitrary 3-manifolds\cite{Turaev,Turaevtop},
the classification of Vassiliev invariants\cite{Bar-Natan} or 
3-manifold invariants\cite{CareyZhang}). Hopefully the present approach
at least has the merit of pointing a way to a more fundamental view of 
the nature of renormalizable quantum field theories (in arbitrary dimensions),
which may be explored in other contexts.\\
\textbf{Note added in proof:} It should be noted that, from the general 
$p$ loop result quoted, the interpretation of the third example in 
Figure \ref{equivexx} would be expected to involve the $(2,5)$ torus knot 
(associated with $\zeta(5)$).   
   
\subsection*{Acknowledgements}
The authors would like to thank Dirk Kreimer for his insistence, during
the course of his work in the theory group in 
Tasmania, that there should be a solid basis for his `knottish' ideas, 
and for comments on a draft of this letter; and 
Hughan Ross for correspondence and alerting us to salient literature.
We also thank Bob Delbourgo for constructive dialogue
and healthy scepticism, other members of the group for
their tolerance, and especially Neville Jones for computing support.
Finally we acknowledge the Australian Research 
Council for a Small Grant project award which made possible the visits of 
ACK and GT.
  

\newpage

\begin{thebibliography}{99}

\bibitem{Kreimer1}
Dirk Kreimer, \emph{Knots and Divergences},
 Phys Lett \textbf{B354} (1995) 117-124
\bibitem{Kreimer2}
Dirk Kreimer, \emph{Renormalization and Knot Theory}, preprint UTAS-PHYS-94-25
(hep-th/9412045), to appear in J Knot Theory and its Ramifications
\bibitem{Kreimer3}
Dirk Kreimer, \emph{Feynman Diagram Calculations - From finite integral
representations to knotted infinities},
in Proc IAHENP95 Conference (Pisa 1995),
\emph{New Computing Techniques in Physics Research IV},
Ed B Denby, D Perret-Gallix, (World Scientific)
\bibitem{KreimerB1}
D J  Broadhurst, D Kreimer, 
\emph{Knots and numbers in $\phi^4$ theory to 
7 loops and beyond}, Int J Mod Phys \textbf{C6} (1995) 519
\bibitem{Alvarez}
M Alvarez and J M F Lambastida,
\emph{Numerical knot invariants of finite type from Chern-Simons perturbation 
theory}, Nucl Phys \textbf{B433} (1995) 555-596
\bibitem{Bar-Natan}
Dror Bar-Natan, \emph{On the Vassiliev Knot Invariants},
Topology \textbf{34},2 (1995) 423-72
\bibitem{Birman}
Joan S Birman, \emph{New Points of View in Knot Theory},
Bulletin (New Series) of the American Mathematical Society \textbf{28},2
(1993) 253-287
\bibitem{BPHZ}
See for example Claude Itzykson and Jean-Bernard Zuber,
\emph{Quantum Field Theory}, New York: McGraw-Hill (1980) Chap 8
\bibitem{Kauffman}
Louis H Kauffman, \emph{Knots and physics}, Singapore: World
Scientific (1991)
\bibitem{BDK}
D J Broadhurst, R Delbourgo, D Kreimer,
\emph{Unknotting the polarized vacuum of quenched QED},
preprint OUT-4102-60, UTAS-PHYS-95-40, MZ-TH/95-22 (hep-ph/9509296)
\bibitem{Turaev}
V G Turaev, \emph{Conway and Kauffman modules of a solid torus},
Zap Nauch Sem LOMI \textbf{167} (1988) 79-89
(English translation: \emph{J Soviet Math})
\bibitem{Turaevtop}
V G Turaev, \emph{Skein quantization of Poisson algebras of loops on surfaces},
Ann scient \'{E}c Norm Sup \textbf{24} (1991) 635-704
\bibitem{CareyZhang}
R B Zhang and A L Carey,
\emph{Quantum groups at odd roots of unity and topological invariants of 
3-manifolds}, preprint, Dept of Pure Mathenatics, University of Adelaide
(1995) 
\end{thebibliography}
\end{document}